\documentclass[%
 aip,
 amsmath,amssymb,
 reprint,%
]{revtex4-1}

\usepackage{graphicx}
\usepackage{dcolumn}
\usepackage{bm}
\usepackage{color}
\usepackage[utf8]{inputenc}
\usepackage[T1]{fontenc}
\usepackage{mathptmx}
\usepackage{etoolbox}

\makeatletter
\def\@email#1#2{%
 \endgroup
 \patchcmd{\titleblock@produce}
  {\frontmatter@RRAPformat}
  {\frontmatter@RRAPformat{\produce@RRAP{*#1\href{mailto:#2}{#2}}}\frontmatter@RRAPformat}
  {}{}
}
\makeatother

\begin{document}


\title{Electronic-Structure Control of Nuclear Excitation by Electron Capture in Finite-Density $^{229}$Th}

\author{Yang-Yang Xu}
\affiliation{College of Science, National University of Defense Technology, Changsha 410073, China}

\author{Jin-Tao Qi}
\altaffiliation{\raggedright Authors to whom correspondence should be addressed: qijintao@sztu.edu.cn, lixinyan\_nudt@nudt.edu.cn and taiwu.huang@sztu.edu.cn.}
\affiliation{Shenzhen Key Laboratory of Ultraintense Laser and Advanced Material Technology, Center for Intense Laser Application Technology, and College of Engineering Physics, Shenzhen Technology University, Shenzhen 518118, China}

\author{Qiong Xiao}
\affiliation{College of Science, National University of Defense Technology, Changsha 410073, China}

\author{Jun-Hao Cheng}
\affiliation{Southwestern Institute of Physics, Chengdu 610003, China }

\author{Xin-Yan Li}
\altaffiliation{\raggedright Authors to whom correspondence should be addressed: qijintao@sztu.edu.cn, lixinyan\_nudt@nudt.edu.cn and taiwu.huang@sztu.edu.cn.}
\affiliation{College of Science, National University of Defense Technology, Changsha 410073, China}

\author{Tai-Wu Huang}
\altaffiliation{\raggedright Authors to whom correspondence should be addressed: qijintao@sztu.edu.cn, lixinyan\_nudt@nudt.edu.cn and taiwu.huang@sztu.edu.cn.}
\affiliation{Shenzhen Key Laboratory of Ultraintense Laser and Advanced Material Technology, Center for Intense Laser Application Technology, and College of Engineering Physics, Shenzhen Technology University, Shenzhen 518118, China}

\author{Tong-Pu Yu}
\affiliation{College of Science, National University of Defense Technology, Changsha 410073, China}

\date{\today}

\begin{abstract}
	Nuclear excitation by electron capture (NEEC) provides a unique pathway for coupling electronic and nuclear dynamics, but its description in dense matter commonly relies on electronic structures of isolated ions. Here we show how dense environments reshape the available NEEC capture channels. Using a finite-temperature average-atom model, we assess channel availability by jointly considering electronic localization, resonance energy matching, and vacancy availability. Near solid density, the shallow $6p$ states remain sufficiently localized to support resonant electron capture that drives the 8.356-eV isomeric transition in $^{229}$Th, whereas higher valence-like states merge into the continuum and no longer constitute localized capture channels. Calculations at different temperatures and densities reveal distinct windows of channel availability arising from the interplay among pressure-induced delocalization, shifts in resonance energy, and vacancy formation. Coupling the reconstructed channels to particle-in-cell simulations of laser-driven $^{229}$Th further shows that electronic structure at finite density can substantially alter the predicted cumulative NEEC yield. These results demonstrate how the electronic environment governs resonant capture pathways, highlighting its essential role in nuclear excitation driven by electrons in dense matter.
\end{abstract}

\maketitle

\makeatletter
\renewenvironment{widetext}{%
	\par\ignorespaces
	\onecolumngrid
	\vskip10\p@
	\prep@math@patch
}{%
	\par
	\vskip8.5\p@
	\twocolumngrid
	\global\@ignoretrue
	\@endpetrue
}
\makeatother

\section{Introduction}
The low-lying isomeric state of $^{229}\mathrm{Th}$ has attracted sustained interest because its excitation energy, approximately $8.356~\mathrm{eV}$ above the nuclear ground state, lies in the vacuum ultraviolet range. \cite{Beeks2021} As the lowest known nuclear excited state, this transition connects nuclear structure with electronic energies on the atomic scale and provides a unique platform for precision spectroscopy and nuclear optical clock development. \cite{refId0,PhysRevLett.106.223001} Recent progress in direct laser excitation of the $^{229}\mathrm{Th}$ nucleus has further demonstrated optical access to this nuclear transition. \cite{PhysRevLett.132.182501,PhysRevLett.133.013201,Zhang2024,Xiao2026LaserReview} These developments have also renewed interest in complementary electron-mediated routes, \cite{Fu2022MRE} in which the nuclear transition is driven indirectly through its coupling to atomic or continuum electrons. Examples include electronic bridge excitation (EB), \cite{PhysRevA.81.042516,PhysRevLett.105.182501,Bilous_2018,PhysRevC.100.044306,PhysRevC.102.024604,PhysRevC.106.064608,PhysRevLett.124.192502,PhysRevLett.125.032501} nuclear excitation by electron transition (NEET), \cite{Dzyublik2011,PhysRevC.88.054616,PhysRevC.74.031301,10.1143/PTP.49.1574,Fujioka_1985,TKALYA1992209,PhysRevLett.85.1831,SAKABE20051} nuclear excitation by electron capture (NEEC), \cite{GOLDANSKII1976393,CUE198925,PhysRevC.47.323,PhysRevLett.112.082501,chiara2018isomer,PhysRevLett.122.212501,PhysRevLett.128.242502} nuclear excitation by inelastic electron scattering (NEIES), \cite{PhysRevLett.124.242501,CPC10.1088/1674-1137/ac9f0a,PhysRevC.106.044604,PhysRevC.106.064604,PhysRev.92.978,PhysRevC.110.064621,s6r5-5y5v} and laser-driven electron recollision. \cite{PhysRevLett.127.052501,PhysRevC.106.024606}

Among these electron-mediated mechanisms, NEEC is particularly sensitive to the electronic structure. \cite{PhysRevA.73.012715,Palffy2010} In NEEC, a continuum electron is captured into a bound electronic vacancy, and the binding energy of the capture state together with the incident electron kinetic energy is transferred resonantly to the nucleus. The resonance condition therefore depends directly on the binding energy and availability of the final capture orbital. In calculations for isolated ions, one usually obtains discrete atomic orbitals for a given charge state and then identifies capture channels with positive resonance energies. For low-energy nuclear transitions such as the $^{229}\mathrm{Th}$ isomer, this procedure tends to select weakly bound outer orbitals and high-lying Rydberg-like states, because their small binding energies can leave a positive resonant electron energy. \cite{Xu2026,82cg-pkxb} This picture is appropriate under conditions typical of dilute ion beams, storage rings, or an electron beam ion trap (EBIT), where each ion can be approximated as an isolated atomic system. \cite{Wang2023EBITNEEC,Yang2024StorageRingNEEC}

Previous studies have investigated NEEC in laser-generated plasmas, with particular emphasis on the roles of electron distributions and charge-state populations in determining nuclear excitation yields. \cite{PhysRevC.59.2462,PhysRevE.97.063205,PhysRevLett.130.112501,PhysRevC.110.L051601,Ma2024} Here we focus on a complementary effect that becomes important at finite density: the surrounding medium can modify the electronic states into which resonant capture occurs. Pressure ionization and band formation may delocalize weakly bound outer states, \cite{PhysRevB.20.4981,More1985} while the binding energies and occupations of the remaining localized states can shift substantially from their isolated-ion values. \cite{Ciricosta2016,1966ApJ...144.1203S} The dense environment can therefore alter not only the electron source for NEEC, but also which capture channels remain physically accessible. These effects are especially important for the 8.356-eV isomeric transition of $^{229}\mathrm{Th}$, for which eV-scale changes in electronic binding energies can strongly modify low-energy NEEC resonances.

\begin{figure}[h]
	\centering
	\includegraphics[width=0.47\textwidth]{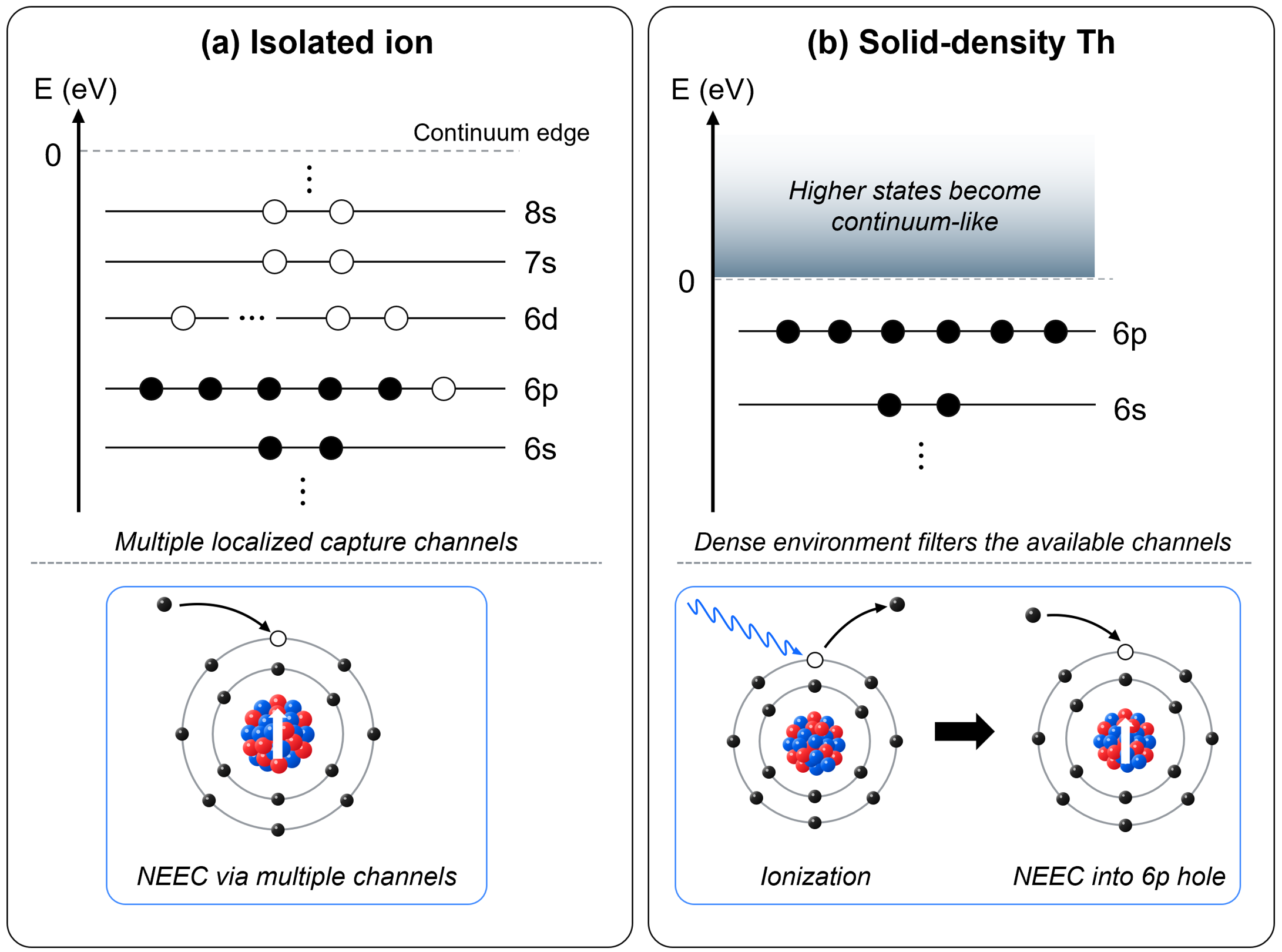}
	\caption{Schematic comparison of NEEC channel availability between the isolated-ion approximation and thorium at finite density. (a) In the isolated-ion picture, several discrete bound orbitals can serve as localized NEEC capture channels, such as $6p$, $7s$, and $8s$ vacancies. (b) In the dense environment, higher-lying states become continuum-like and no longer retain localized character relevant to NEEC capture, while the shallow $6p$ state remains as a bound-like capture state. Since the $6p$ shell is nearly filled in cold thorium at solid density, vacancy creation is required before the surviving $6p$ channel can contribute to NEEC.}
	\label{fig:concept}
\end{figure}

In this article, we investigate the role of finite-density electronic structure in controlling NEEC capture pathways in $^{229}\mathrm{Th}$. Using a finite-temperature Kohn--Sham average-atom model implemented in the open-source atoMEC code, \cite{PhysRev.75.1561,Callow2022AtoMEC,PhysRevResearch.5.013049,Hou2021MRE,Ovechkin2022MRE,HOU201721,PhysRevResearch.3.023026,Li2025MRE} we determine the localization, binding energies, and occupations of electronic states in dense thorium. We find that near solid density the shallow $6p$ states remain localized and can support low-energy resonant capture, whereas higher valence-like states become continuum-like and no longer constitute localized NEEC channels, as illustrated in Fig.~\ref{fig:concept}. Calculations at different temperatures and densities further reveal that channel availability is jointly governed by electronic localization, resonance energy matching, and vacancy formation. Finally, by combining these capture channels with the output of particle-in-cell (PIC) simulations in NEEC postprocessing, we demonstrate how electronic structure at finite density reshapes the available capture pathways and thereby modifies the predicted nuclear excitation yield in laser-driven $^{229}\mathrm{Th}$.

\section{Theoretical Framework}
\label{sec:theory}

In dense matter, a NEEC channel exists only if the final electronic state remains localized, lies within the energy window set by the nuclear transition, and contains an accessible vacancy. Here we use the electronic structure at finite density to determine the physical accessibility of NEEC capture pathways in dense $^{229}\mathrm{Th}$. The electronic structure is calculated with atoMEC, an open-source implementation of the finite-temperature Kohn-Sham average-atom model. \cite{PhysRev.75.1561,Callow2022AtoMEC,PhysRevResearch.5.013049} For a thorium mass density $\rho$, the ion number density is $n_i=\rho/M_{\rm Th}$, where $M_{\rm Th}$ is the mass of one thorium atom. In the average-atom description, each thorium nucleus is assigned one Wigner-Seitz ion sphere whose volume equals the volume per ion, $V_{\rm IS}=1/n_i$. The corresponding ion sphere radius is
\begin{equation}
	R_{\rm IS}=\left(\frac{3}{4\pi n_i}\right)^{1/3}.
\end{equation}
At a given electronic temperature $T_e$, one thorium nucleus and its electrons are embedded in this ion sphere, and the surrounding medium is represented through the boundary condition at $R_{\rm IS}$. In the spherically symmetric average-atom model, the one-electron states are obtained from the finite-temperature Kohn-Sham equation, \cite{PhysRev.140.A1133,PhysRev.137.A1441} written schematically in atomic units as
\begin{equation}
	\left[-\frac{1}{2}\nabla^2+V_{\rm KS}[n](r)\right]\psi_\nu(\mathbf r) =\epsilon_\nu\psi_\nu(\mathbf r),
\end{equation}
where $r=|\mathbf r|$ is the radial coordinate, $\psi_\nu(\mathbf r)$ is the Kohn-Sham orbital, and $\epsilon_\nu$ is the corresponding Kohn-Sham eigenvalue. The index $\nu$ collectively labels the orbital quantum numbers and the integration points for the bands in the average-atom calculation. The effective potential $V_{\rm KS}[n](r)$ is a functional of the self-consistent electron density $n(r)$ and contains the electron-nucleus attraction, the Hartree potential, and the exchange-correlation potential,
\begin{equation}
	V_{\rm KS}[n](r) = V_{\rm en}(r)+V_{\rm H}[n](r)+V_{\rm xc}[n](r).
\end{equation}
The electron density $n(r)$ is constructed self-consistently from the Kohn-Sham orbitals with Fermi-Dirac occupations. In the present calculations, neutral thorium average atoms are considered, so the total electron number in the ion sphere is fixed to $Z=90$. The Kohn-Sham potential is shifted by its value at the ion sphere boundary, so that the continuum edge is taken as the working zero of energy. \cite{More1985,PhysRevB.20.4981}

For a given $nl$ band, where $n$ and $l$ denote the principal and orbital angular momentum quantum numbers, respectively, the average-atom output provides a set of integration points for the band, labeled by $k$. \cite{Callow2022AtoMEC,PhysRevResearch.5.013049} Each point has a Kohn-Sham energy $\epsilon_{nlk}$, a Fermi-Dirac occupation $f_{nlk}$, a factor $D_{nlk}$ for the density of states, and an integration weight $w_k$. For an unpolarized spin population, the degeneracy is $g_l=2(2l+1)$, and the state weight at each integration point is $\Omega_{nlk}=D_{nlk}g_lw_k$. The occupation of the $nl$ band is then $N_{nl}=\sum_k f_{nlk}\Omega_{nlk}$. The lower and upper edges of the band are $\epsilon_{nl}^{\min}=\min_k\epsilon_{nlk}$ and $\epsilon_{nl}^{\max}=\max_k\epsilon_{nlk}$, respectively, and the weighted mean band energy is
\begin{equation}
	\bar{\epsilon}_{nl}= \frac{\sum_k \epsilon_{nlk}\Omega_{nlk}} {\sum_k \Omega_{nlk}} .
\end{equation}
Here $\bar{\epsilon}_{nl}$ characterizes the band position, while $N_{nl}$ characterizes its thermal occupation.

The average-atom calculation is resolved in $n$ and $l$, but not in the relativistic fine-structure quantum number $j$. Thus an $nl$ band includes the corresponding fine-structure components, such as $6p_{1/2}$ and $6p_{3/2}$ for the $6p$ band. When comparing with NEEC calculations that resolve the $j$ components for isolated ions, we use the binding energy averaged over the $nl$ band as an approximate environmental modification for the corresponding fine-structure capture states.

The position of a band relative to the continuum edge is used to classify the corresponding capture state. A band is treated as bound-like when $\epsilon_{nl}^{\max}<0$, continuum-like when $\epsilon_{nl}^{\min}>0$, and near-edge when $\epsilon_{nl}^{\min}<0<\epsilon_{nl}^{\max}$. Near-edge bands are not counted as robust localized capture states unless stated otherwise. For a bound-like band, the binding energy window is defined by $B_{nl}^{\min}=-\epsilon_{nl}^{\max}$, $B_{nl}^{\rm mean}=-\bar{\epsilon}_{nl}$, and $B_{nl}^{\max}=-\epsilon_{nl}^{\min}$.

For a bound-like capture channel \(\alpha\), the resonance energy is
\begin{equation}
	E_r^\alpha=E_{\rm nuc}-B_{\alpha}^{\rm env},
\end{equation}
where \(E_{\rm nuc}=8.356~\mathrm{eV}\) is the nuclear transition energy of the \(^{229}\mathrm{Th}\) isomer and \(B_{\alpha}^{\rm env}\) is the environmental binding energy of the capture state. Unless otherwise stated, \(B_{\alpha}^{\rm env}\) is taken as the weighted mean binding energy \(B_{nl}^{\rm mean}\) of the corresponding bound-like band. The band edges determine whether the corresponding channel is open, partial, or closed. Specifically, the channel is classified as open if $B_{nl}^{\max}<E_{\rm nuc}$, partial if $B_{nl}^{\min}<E_{\rm nuc}<B_{nl}^{\max}$, and closed if $B_{nl}^{\min}>E_{\rm nuc}$.

For the comparison of individual channels below, the NEEC resonance strength is used to characterize each contributing capture channel. It is defined as the integrated resonant cross section, \cite{PhysRevA.73.012715,Palffy2010,Zhang2023ElectronicProcessesTh229}
\begin{equation}
	S_\alpha(E_r^\alpha)=\int \sigma_\alpha(E_r^\alpha,E_e)\,dE_e,
\end{equation}
where $E_e$ is the kinetic energy of the continuum electron and $\sigma_\alpha(E_r^\alpha,E_e)$ is the resonant NEEC cross section for channel $\alpha$ at the resonance energy $E_r^\alpha$.

To quantify how resonance placement affects the overlap with continuum electrons, we evaluate the NEEC rate by integrating the relevant resonant contributions over a normalized electron spectrum $f(E_e)$. For charge state $q$, the NEEC excitation rate per ion for the specified electronic configuration is
\begin{equation}
	\lambda_q(n_e,f) = n_e \sum_{\alpha\in q} S_\alpha(E_r^\alpha)v(E_r^\alpha)f(E_r^\alpha), \label{eq:lambda_q}
\end{equation}
where $n_e$ is the density of continuum electrons, $v(E_e)$ is the nonrelativistic electron speed, and $f(E_e)$ is normalized as $\int f(E_e)dE_e=1$. The resonance strength \(S_\alpha(E_r^\alpha)\) is converted from \(\mathrm{barn\,eV}\) to \(\mathrm{m^2\,eV}\), while \(f(E_e)\) has units of \(\mathrm{eV}^{-1}\), so that \(\lambda_q\) is given in \(\mathrm{s}^{-1}\).

\section{Results and Discussion}
\label{sec:results}

\begin{figure}[h]
	\centering
	\includegraphics[width=0.9\linewidth]{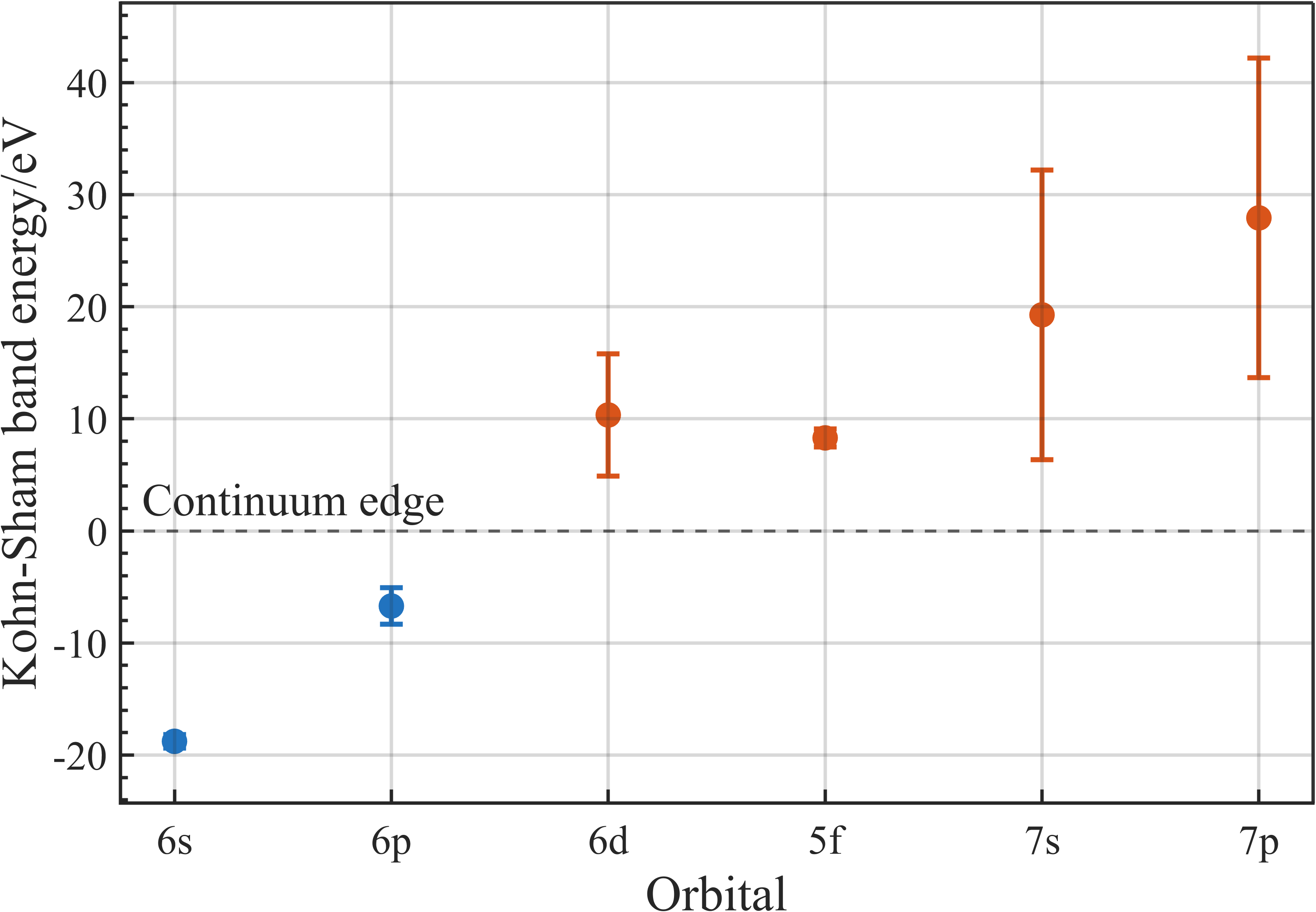}
	\caption{Kohn-Sham band positions of selected shallow electronic states in metallic thorium at $\rho=\rho_0$ and $T_e=0.1~\mathrm{eV}$. The vertical bars show the energy windows of the bands, and the filled circles show the weighted mean band energies. The dashed horizontal line marks the continuum edge.}
	\label{fig:baseline_bands}
\end{figure}

We first consider metallic thorium at the reference density $\rho_0=11.7~\mathrm{g/cm^3}$ and at a low electronic temperature $T_e=0.1~\mathrm{eV}$. This case serves as the baseline for identifying which shallow electronic states remain available as localized capture states under these conditions. Figure~\ref{fig:baseline_bands} shows the Kohn-Sham band positions of selected states, and the corresponding numerical values are listed in Table~\ref{tab:baseline_bands}. All Kohn-Sham energies are measured relative to the continuum edge.

The baseline calculation shows that finite density strongly restricts the available low-energy capture states. Among the selected shallow states, only the $6s$ and $6p$ bands remain below the continuum edge. The $6s$ band is bound-like but closed for the $8.356~\mathrm{eV}$ transition because its binding energy is too large. By contrast, the $6p$ band remains bound-like and has a binding energy window of approximately $5.07$--$8.34~\mathrm{eV}$, just below $E_{\rm nuc}$. Thus, among the selected shallow states, $6p$ is the only localized band that satisfies the NEEC energy criterion at metallic density.

\begin{table}[h]
	\caption{Baseline band quantities for metallic thorium at $\rho=\rho_0$ and $T_e=0.1~\mathrm{eV}$. The Kohn-Sham band energies are measured relative to the continuum edge. The status codes are BO: bound-like and open, BC: bound-like but closed, and CL: continuum-like.}
	\label{tab:baseline_bands}
	\begin{ruledtabular}
		\begin{tabular}{ccccc}
			Orbital & $\epsilon_{nl}^{\min}$--$\epsilon_{nl}^{\max}/\mathrm{eV}$ & $\bar{\epsilon}_{nl}/\mathrm{eV}$ & $N_{nl}$ & Status \\
			\hline
			$6s$ & $-19.38$--$-18.17$ & $-18.78$ & $1.99$ & BC \\
			$6p$ & $-8.34$--$-5.07$ & $-6.71$ & $5.96$ & BO \\
			$6d$ & $4.88$--$15.80$ & $10.34$ & $2.09$ & CL \\
			$5f$ & $7.48$--$9.11$ & $8.30$ & $1.93$ & CL \\
			$7s$ & $6.34$--$32.21$ & $19.27$ & $0.03$ & CL \\
			$7p$ & $13.66$--$42.19$ & $27.92$ & $\simeq0$ & CL \\
		\end{tabular}
	\end{ruledtabular}
\end{table}

The occupation gives a further constraint. Although the $6p$ channel is energetically open, the shell is nearly filled in the cold metallic baseline, with $N_{6p}=5.96$ out of the maximum capacity of six electrons. The corresponding equilibrium hole number is only $N_{h,6p}\simeq0.04$. The cold metallic system therefore provides a $6p$ channel that satisfies the resonance condition, but only a very small equilibrium vacancy population. It is worth noting that bands with positive energies should not be interpreted as ordinary valence orbitals of isolated atoms. In particular, the approximately four electrons projected onto the continuum bands with $6d$ and $5f$ character represent a continuum-like or conduction-like population in metallic thorium, rather than an isolated atomic configuration. Consequently, continuum-like electrons may contribute to the electron source spectrum, but they do not define localized NEEC capture vacancies.

\begin{figure}[h]
	\centering
	\includegraphics[width=0.9\linewidth]{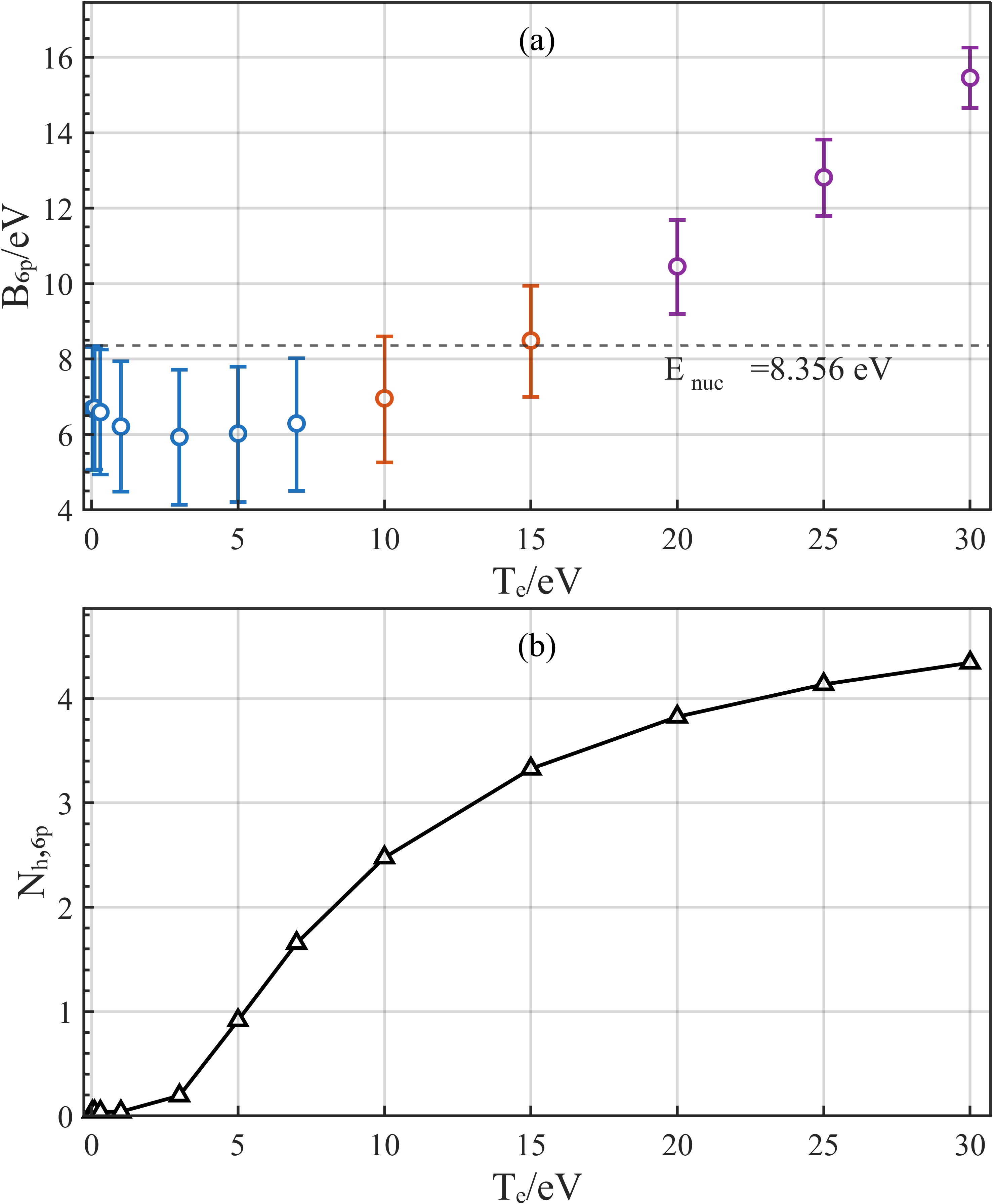}
	\caption{Temperature dependence of the $6p$ state relevant to NEEC at metallic density $\rho=\rho_0$. (a) $6p$ binding energy window as a function of electronic temperature. The vertical bars show $B_{6p}^{\min}$--$B_{6p}^{\max}$, and the open circles show the weighted mean binding energy. Blue, orange, and purple symbols denote open, partial, and closed channel status, respectively. The dashed horizontal line marks $E_{\rm nuc}=8.356~\mathrm{eV}$. (b) Corresponding $6p$ hole number, $N_{h,6p}=6-N_{6p}$, shown by open triangles.}
	\label{fig:Te_6p}
\end{figure}

We next examine how temperature modifies the $6p$ electronic state that remains bound-like at metallic density. Figure~\ref{fig:Te_6p} shows the $6p$ binding energy window and the corresponding hole number $N_{h,6p}=6-N_{6p}$ as functions of $T_e$ at fixed $\rho=\rho_0$.

Starting from the cold metallic baseline, increasing $T_e$ produces two competing effects. Thermal depopulation increases the number of $6p$ holes, while the mean binding energy initially decreases slightly and then increases at higher temperatures. Around $T_e=5$--$7~\mathrm{eV}$, the $6p$ band remains energetically open while the hole population is markedly increased. This regime is therefore more favorable than the cold metallic baseline because both the resonance condition and vacancy availability are improved. At higher temperatures, vacancy formation alone is no longer sufficient. The $6p$ binding energy window crosses $E_{\rm nuc}$ around $T_e=10$--$15~\mathrm{eV}$ and moves above the nuclear transition energy at still higher temperatures. The channel then becomes only partially open or energetically closed despite the larger number of holes. The temperature scan therefore shows that vacancy availability and resonance energy matching must be satisfied simultaneously; increasing temperature does not monotonically improve the conditions for NEEC into $6p$.

\begin{figure}[h]
	\centering
	\includegraphics[width=0.9\linewidth]{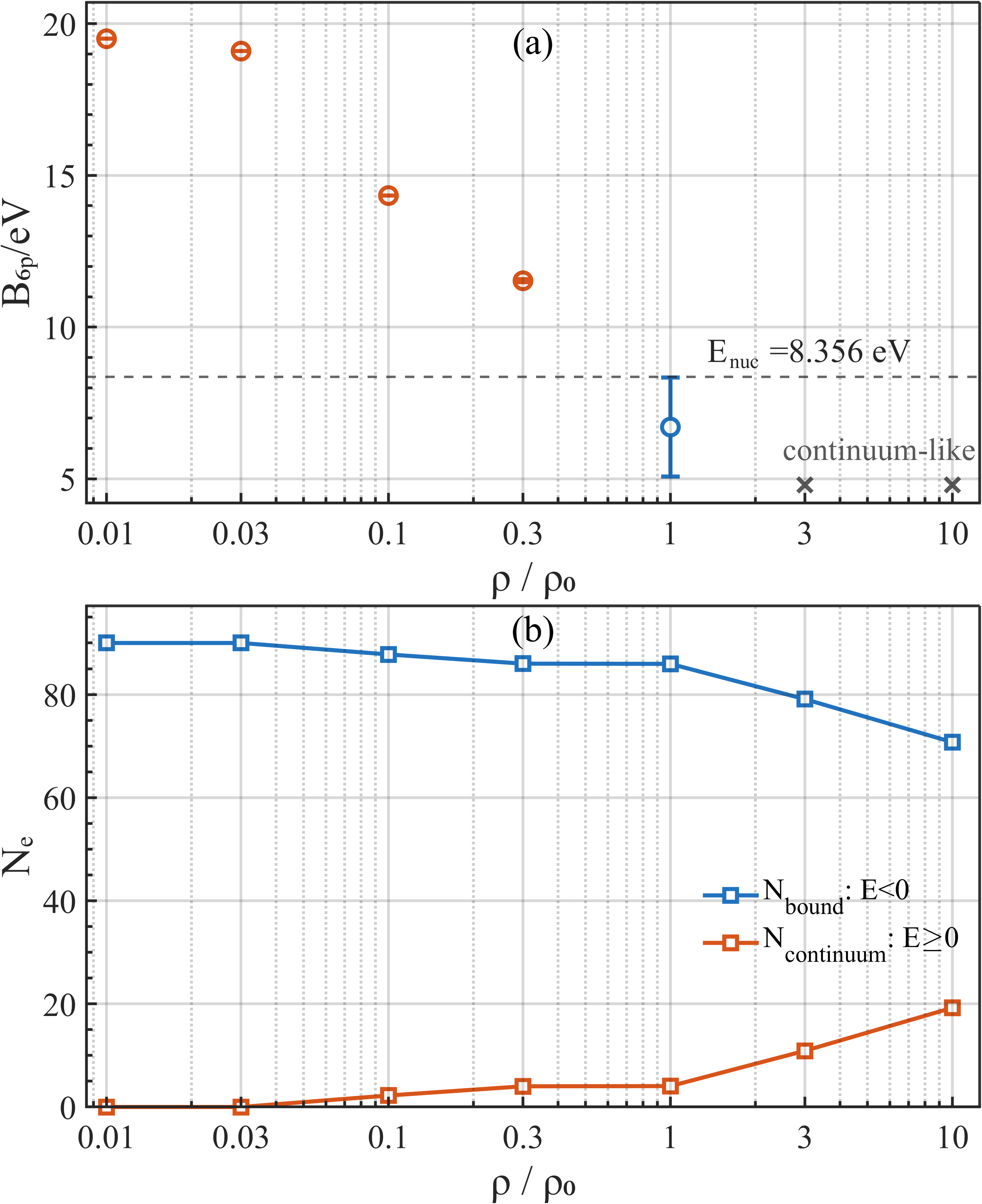}
	\caption{Electronic structure as a function of normalized thorium mass density $\rho/\rho_0$ at $T_e=0.1~\mathrm{eV}$, where $\rho_0=11.7~\mathrm{g/cm^3}$ is the metallic density. (a) Binding energy window of the $6p$ band. The vertical bars denote $B_{6p}^{\min}$--$B_{6p}^{\max}$, and the open circles denote the weighted mean binding energy. Orange and blue symbols indicate closed and open $6p$ channels, respectively. The dashed horizontal line marks $E_{\rm nuc}=8.356~\mathrm{eV}$. The gray crosses indicate densities where the $6p$ band becomes continuum-like and no longer represents a localized capture state for NEEC. (b) Numbers of bound-like and continuum-like electrons per average-atom ion sphere, shown by open square markers.}
	\label{fig:density_6p}
\end{figure}

The temperature scan above shows how vacancy availability and energy matching compete at fixed metallic density. A complementary question is how changing the density modifies the orbital character itself. Figure~\ref{fig:density_6p} therefore shows the density evolution of the $6p$ binding energy window and the redistribution of electrons between bound-like and continuum-like states at fixed $T_e=0.1~\mathrm{eV}$.

At low densities, $\rho/\rho_0=0.01$--$0.3$, the $6p$ band remains bound-like but its binding energy window lies above $E_{\rm nuc}$. These cases therefore retain localized $6p$ states, but the low-energy NEEC channel is closed because the capture binding energy exceeds the nuclear transition energy. Near the metallic density, the $6p$ binding energy window shifts to approximately $5.07$--$8.34~\mathrm{eV}$, just below $E_{\rm nuc}$. The $6p$ channel is therefore both bound-like and energetically open at $\rho=\rho_0$. This identifies a density window near metallic density in which the resonance condition is satisfied: the localized $6p$ capture state survives and its binding energy falls into the low-energy NEEC range. At compressed densities, e.g., $\rho/\rho_0=3$ and $10$, the $6p$ band becomes continuum-like. Assigning a localized NEEC binding energy to this band is then no longer meaningful because the bound final state required for NEEC is absent. This behavior is also reflected in Fig.~\ref{fig:density_6p}(b), where increasing density transfers electrons from bound-like bands to continuum-like states. Thus, the density affects the $6p$ channel in two ways: it shifts the binding energy while the band remains bound-like, and at sufficiently high density it removes the localized $6p$ capture channel altogether.

The present analysis focuses on the density region around solid density relevant to the target conditions considered here. In this regime, although other shallow states may appear under different thermodynamic conditions, the localized $6p$ state provides the most persistent low-energy capture channel over the broad parameter range investigated. In the low-density limit, the system gradually approaches the isolated-ion regime, where more discrete bound states can become relevant NEEC channels.

To illustrate how electronic structure at finite density modifies the NEEC resonance structure, we compare the isolated-ion $\mathrm{Th}^{5+}$ reference with the calculation at metallic density. For the latter, the capture states and their environmental binding energies are obtained from the average-atom calculation. For the localized $6p$ band, the weighted mean binding energy is used to represent the corresponding capture channel. The comparison assumes that the corresponding localized vacancies are available, allowing the effects of changes in electronic states and resonance shifts to be separated from vacancy production.

\begin{figure}[h]
	\centering
	\includegraphics[width=0.9\linewidth]{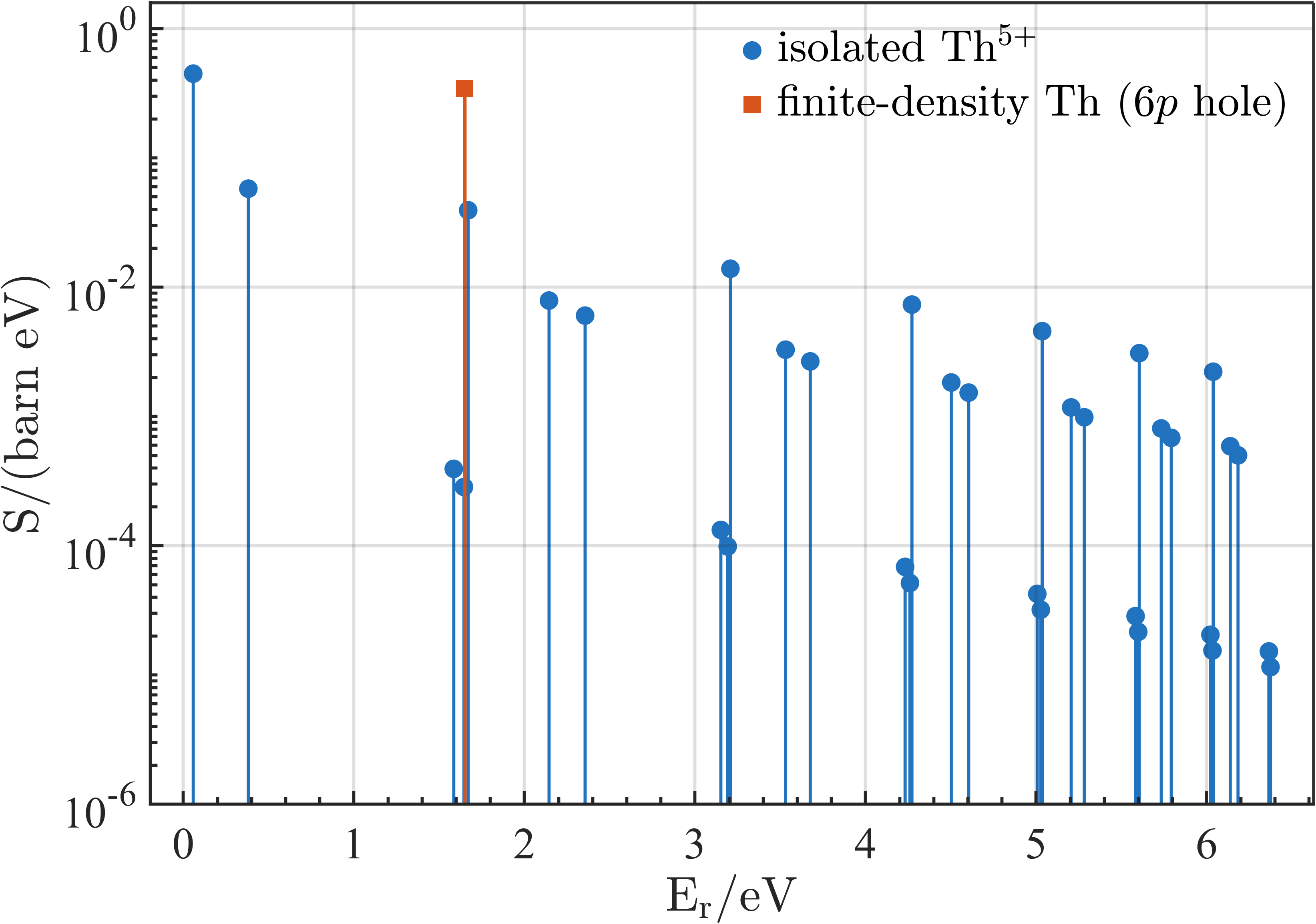}
	\caption{Comparison of NEEC resonance strengths for an isolated $\mathrm{Th}^{5+}$ reference and the representative capture channels obtained from metallic thorium. The points denote the resonance strengths, and the vertical lines indicate the corresponding resonance positions.}
	\label{fig:S_compare}
\end{figure}

Figure~\ref{fig:S_compare} shows that thorium at finite density retains only a limited number of localized resonances compared with the isolated-ion reference. This reduction follows from the channel criteria introduced above: capture orbitals that are bound in the isolated-ion calculation can become continuum-like in the dense environment and lose their contribution as localized NEEC states. However, the dominant resonance strength is not reduced by the same factor as the number of channels. The strongest remaining resonance is comparable in strength to the strongest isolated-ion resonances, because many of the excluded states are weak high-lying channels. Thus, channel multiplicity alone does not determine the dominant NEEC contribution.

Finite density also shifts the resonance positions. According to $E_r^\alpha=E_{\rm nuc}-B_\alpha^{\rm env}$, the environmental binding energy sets the energy of the continuum electron at which each resonance contributes. The excitation rate is therefore controlled not only by the presence of localized channels, but also by how their shifted resonance positions overlap with the electron energy distribution.

We then evaluate the NEEC rates for the isolated-ion and finite-density descriptions using prescribed normalized electron spectra and Eq.~(\ref{eq:lambda_q}). The same density of continuum electrons, \(n_e=5n_i\), is used in both cases, so that the comparison isolates the effects of electronic state availability and resonance placement from changes in electron density. The Gaussian spectrum is \(f_G(E_e)=C^{-1}\exp[-(E_e-E_0)^2/(2\Delta E^2)]\) for \(E_e\geq0\), with \(\Delta E=0.50~\mathrm{eV}\) and \(C\) determined by \(\int_0^\infty f_G(E_e)\,dE_e=1\). The Maxwell-like spectrum is \(f_M(E_e)=(2/\sqrt{\pi})E_e^{1/2}T_{\rm spec}^{-3/2} \exp(-E_e/T_{\rm spec})\), where \(T_{\rm spec}\) is a spectral parameter rather than the average-atom electronic temperature \(T_e\).

\begin{figure}[h]  
	\centering
	\includegraphics[width=0.85\linewidth]{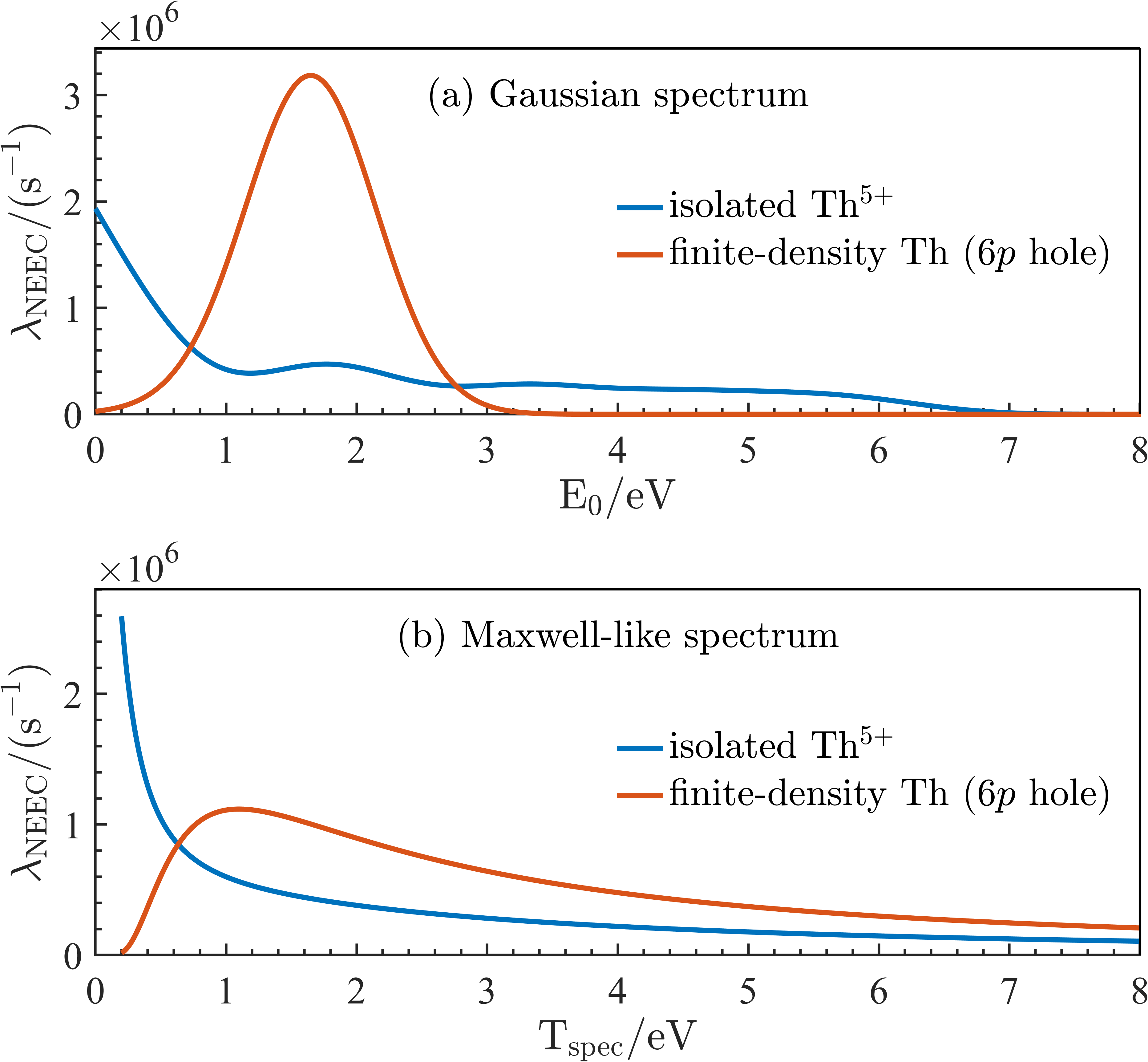}
	\caption{NEEC excitation rates for the isolated $\mathrm{Th}^{5+}$ reference and the finite-density calculation under different electron spectra. The same density of continuum electrons $n_e=5n_i$ is used in both cases. (a) Gaussian electron spectrum with fixed width $\Delta E=0.50~\mathrm{eV}$ as a function of the center energy $E_0$. (b) Maxwell-like electron spectrum as a function of the spectral parameter $T_{\rm spec}$. The rates are given per ion for the specified electronic configurations and corresponding NEEC channels.}
	\label{fig:rate_compare}
\end{figure}

Figure~\ref{fig:rate_compare}(a) shows the result for a Gaussian electron spectrum with fixed width. The finite-density description gives a narrow rate maximum when the spectral center overlaps the shifted \(6p\) resonance. By contrast, the isolated-ion reference produces a broader low-energy response, reflecting contributions from several isolated-ion resonances. For the Maxwell-like spectrum in Fig.~\ref{fig:rate_compare}(b), the isolated-ion rate is enhanced at the lowest \(T_{\rm spec}\) shown because of resonances near threshold, whereas the finite-density rate is favored at intermediate \(T_{\rm spec}\) where the electron distribution overlaps the shifted \(6p\) resonance. Thus, the relative rate is controlled by the spectral overlap with the retained resonance positions, not simply by the number of channels.

The preceding comparison shows how electronic structure at finite density modifies the available resonances and their overlap with electron spectra. We now combine this description with time-dependent plasma conditions obtained from 2D3V PIC simulations. The PIC output provides charge state fractions, electron densities, and electron energy spectra, which are combined with the corresponding NEEC descriptions in postprocessing. We compare a reference case based on an initially neutral target and isolated-ion electronic states with a finite-density description, using the same geometry for the laser and target.

The PIC simulations are performed with EPOCH \cite{Arber2015PPCF} for a thorium slab at solid density, with mass density \(\rho_0=11.7~\mathrm{g/cm^{3}}\). The computational domain is \(0\le x\le 2.0~\mu\mathrm{m}\) and \(-3.0~\mu\mathrm{m}\le y\le 3.0~\mu\mathrm{m}\), resolved by \(1000\times600\) cells. The target is a slab with a thickness of \(100~\mathrm{nm}\) located at \(1.00~\mu\mathrm{m}\le x\le1.10~\mu\mathrm{m}\) and \(-2.20~\mu\mathrm{m}\le y\le2.20~\mu\mathrm{m}\). A Gaussian laser pulse is injected from the \(x_{\min}\) boundary and propagates along the \(+x\) direction. The laser has wavelength \(\lambda_0=0.8~\mu\mathrm{m}\), peak intensity \(I_0=2\times10^{14}~\mathrm{W/cm^{2}}\), temporal profile \(\exp[-(t-t_0)^2/\tau^2]\) with \(t_0=70~\mathrm{fs}\) and \(\tau=25~\mathrm{fs}\), and transverse spot size \(w_0=1.0~\mu\mathrm{m}\). Coulomb collisions, field ionization, and electron-impact ionization are included. \cite{Arber2015PPCF,Sentoku2008JCP,Augst1991JOSAB,Perez2012PoP,Medina2026CPC} The NEEC postprocessing is evaluated over the full target thickness in \(x\) and the central laser-heated region \(|y|\le1.00~\mu\mathrm{m}\).

We consider two cases when calculating NEEC from the PIC output. In the reference case, the target is initialized as neutral thorium; free electrons and charge states are generated by laser-driven ionization, and the NEEC postprocessing uses isolated-ion NEEC channels for the charge states produced in the PIC simulation. In the finite-density case, the target is initialized as an electrically neutral \(\mathrm{Th}^{4+}+4e^-\) plasma. This initialization follows from the average-atom result at metallic density, where the shallow localized manifold is mainly restricted to the \(6s\) and \(6p\) bands and approximately four electrons are projected onto continuum-like states. The \(\mathrm{Th}^{4+}+4e^-\) initialization therefore represents the population of continuum electrons at finite density, rather than a gas of isolated \(\mathrm{Th}^{4+}\) ions. It should be noted that all NEEC calculations assume ground electronic configurations for the initial ions, neglecting contributions from electronically excited configurations.

The ionization energy table used by the modules for field ionization and collisional ionization is also updated to account for ionization potential depression (IPD). \cite{1966ApJ...144.1203S} This makes the charge state evolution consistent with the electronic structure used here. For NEEC postprocessing in this case, the PIC charge state \(q\) is used to identify the corresponding vacancy configuration derived from the finite-density reference state. Starting from the electrically neutral \(\mathrm{Th}^{4+}+4e^-\) state, removal of the localized \(6p\) electrons is mapped sequentially onto the average-atom \(6p\) band at metallic density, across the \(5.07\)--\(8.34~\mathrm{eV}\) binding energy window, and assigned in order to the \(q=5\)--10 configurations. The resonance position for each configuration is then evaluated from the corresponding assigned \(6p\) binding energy, while the resonance strengths are taken from the calculation of NEEC matrix elements for isolated ions with the same vacancy symmetry.

\begin{widetext}
	\begin{center}
		\includegraphics[width=0.8\textwidth]{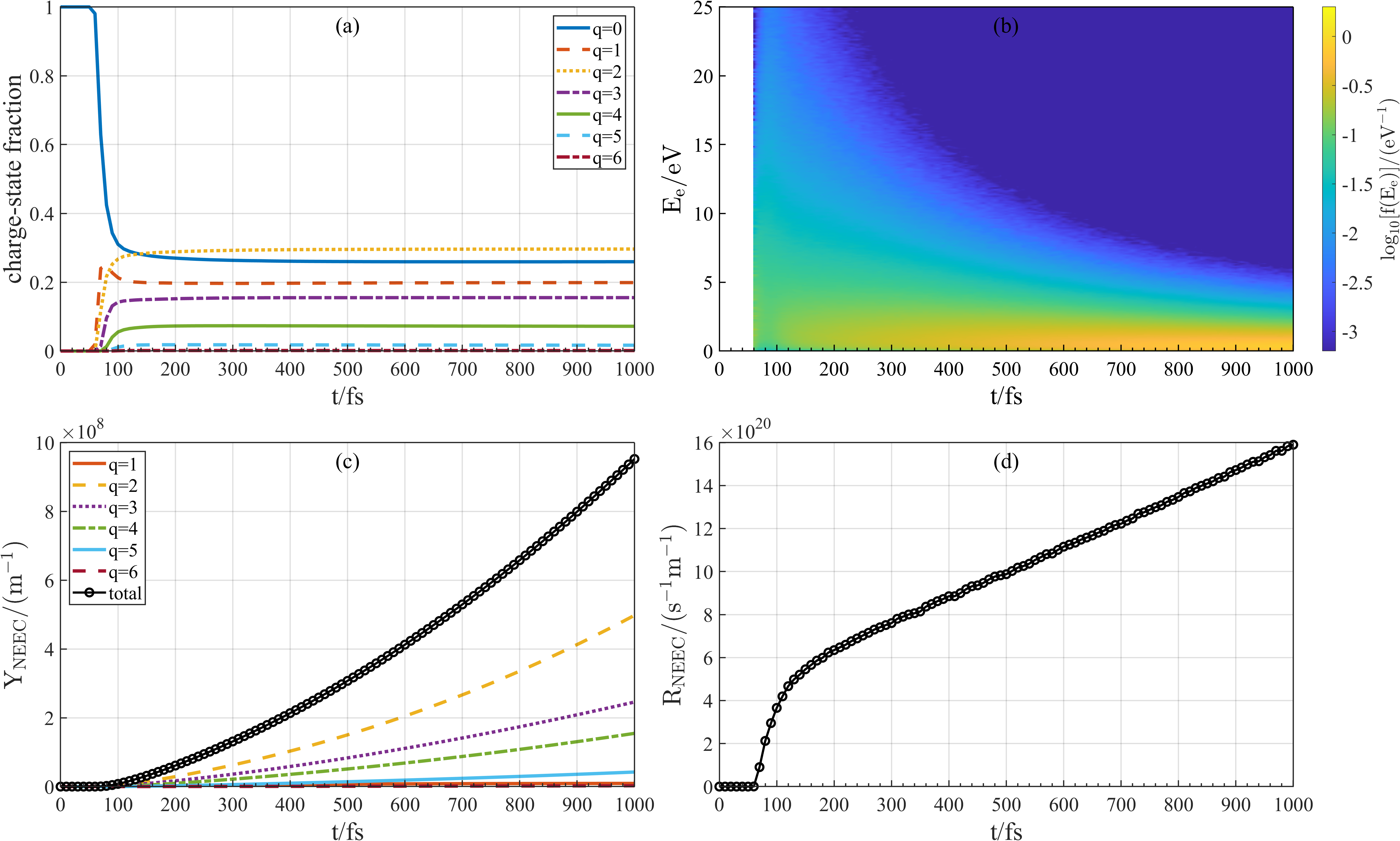}
		\refstepcounter{figure}
		\label{fig:neutral_pic_source}
		\begin{minipage}{0.95\textwidth}
			\small
			FIG.~\thefigure.
			NEEC results obtained from the PIC output for the reference case with an initially neutral target. (a) Charge state fractions in the central laser-heated target region, shown for \(q=0\)--6. (b) Normalized energy distribution of free electrons \(f_t(E_e)\). The blank region at early times indicates that no normalized spectrum of free electrons is available before sufficient laser-driven ionization occurs. (c) Cumulative NEEC yield for each charge state, \(Y_{\rm NEEC}\). (d) Instantaneous NEEC rate, \(R_{\rm NEEC}\). Both quantities are given per unit length perpendicular to the simulation plane.
		\end{minipage}
	\end{center}
\end{widetext}

The NEEC yield is evaluated by folding the evolving plasma profiles from PIC simulations with the rates defined in Eq.~(\ref{eq:lambda_q}), which are given per ion for each charge state. Since the calculation is 2D3V, the reported rate and yield are given per unit length perpendicular to the simulation plane. We denote them by \(R_{\rm NEEC}(t)\) and \(Y_{\rm NEEC}(t)\), respectively:
\begin{equation}
	R_{\rm NEEC}(t)
	=
	\sum_q
	\int_{A_{\rm act}}
	n_q(\mathbf{r},t)
	\lambda_q[n_e(\mathbf{r},t),f_t]
	\,dA .
	\label{eq:pic_rate_2d}
\end{equation}
Here \(A_{\rm act}\) is the selected laser-heated target region, \(n_q(\mathbf{r},t)\) is the density of ions with charge state \(q\), \(n_e(\mathbf{r},t)\) is the density of free electrons, and \(f_t(E_e)\) is the normalized electron energy distribution averaged over \(A_{\rm act}\) at time \(t\), with \(\int f_t(E_e)dE_e=1\). The spatial dependence enters through \(n_q(\mathbf{r},t)\) and \(n_e(\mathbf{r},t)\), while the spectrum averaged over this region sets the overlap with the NEEC resonances. With this definition, \(R_{\rm NEEC}\) has units of \(\mathrm{s^{-1}\,m^{-1}}\). The cumulative yield is
\begin{equation}
	Y_{\rm NEEC}(t)
	=
	\int_0^t
	R_{\rm NEEC}(t')\,dt' ,
	\label{eq:pic_yield_2d}
\end{equation}
with units of \(\mathrm{m^{-1}}\).

Figure~\ref{fig:neutral_pic_source} shows the reference case evaluated with isolated-ion NEEC channels. Since the target is initially neutral, free electrons are generated only after laser-driven ionization begins, at about \(t\simeq60~\mathrm{fs}\). Accordingly, no normalized spectrum of free electrons is available at the earliest times in Fig.~\ref{fig:neutral_pic_source}(b). After ionization, the charge state distribution remains concentrated in low charge states, with sizable populations in \(q=0\)--3. The cumulative NEEC yield is therefore dominated by the corresponding isolated-ion channels and reaches \(Y_{\rm NEEC}\simeq 9.52\times10^{8}~\mathrm{m^{-1}}\) at \(t=1000~\mathrm{fs}\). The instantaneous rate increases gradually after the onset of ionization and remains on the order of \(10^{21}~\mathrm{s^{-1}\,m^{-1}}\) during the later stage of the simulation.

Figure~\ref{fig:fd_pic_source} shows the corresponding finite-density case. Here the initial population of continuum electrons allows a normalized spectrum of free electrons to be defined from the beginning of the simulation. The charge state distribution evolves from the initial \(\mathrm{Th}^{4+}\) plasma toward a mixture mainly composed of \(q=7\) and \(q=8\), with a smaller \(q=9\) population also appearing. Although the \(q=9\) fraction is not the largest, it gives the largest NEEC contribution because its available resonances overlap more favorably with the transient electron spectrum. The instantaneous rate rises rapidly during the early ionization stage, reaches a peak on the order of \(10^{22}~\mathrm{s^{-1}\,m^{-1}}\), and then decreases as the charge state distribution and electron spectrum evolve.

\begin{widetext}
	\begin{center}
		\includegraphics[width=0.8\textwidth]{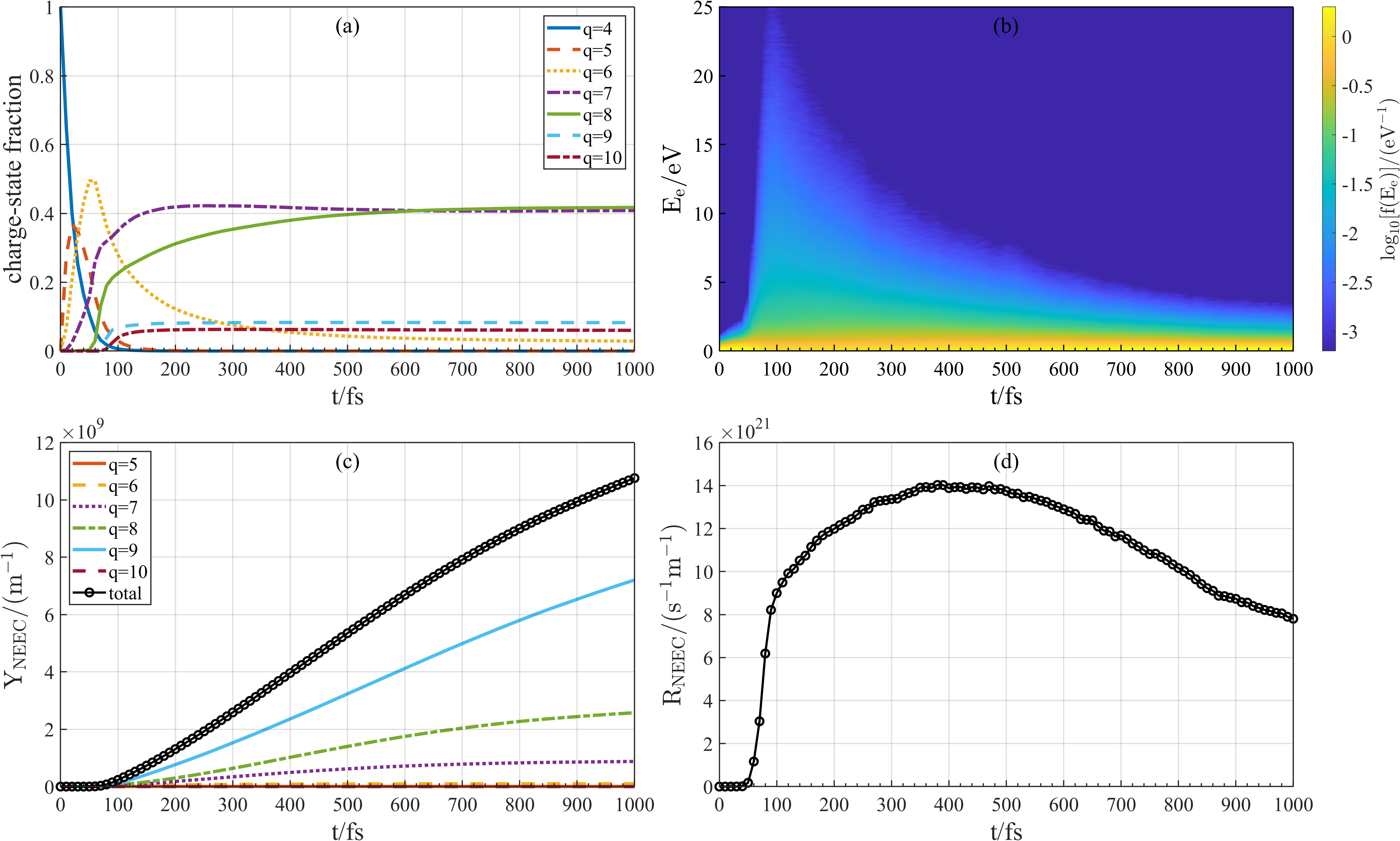}
		\refstepcounter{figure}
		\label{fig:fd_pic_source}
		\begin{minipage}{0.95\textwidth}
			\small
			FIG.~\thefigure.
			NEEC results obtained from the PIC output for the case at finite density. The target is initialized as a \(\mathrm{Th}^{4+}+4e^{-}\) plasma according to the average-atom result at metallic density, and the subsequent ionization uses ionization energies corrected for IPD. (a) Charge state fractions in the central laser-heated target region, shown for \(q=4\)--10. (b) Normalized energy distribution of free electrons \(f_t(E_e)\). (c) Cumulative NEEC yield for each charge state, \(Y_{\rm NEEC}\). (d) Instantaneous NEEC rate, \(R_{\rm NEEC}\). Both quantities are given per unit length perpendicular to the simulation plane.
		\end{minipage}
	\end{center}
\end{widetext}

At \(t=1000~\mathrm{fs}\), the finite-density case reaches \(Y_{\rm NEEC}\simeq 1.08\times10^{10}~\mathrm{m^{-1}}\), about a factor of 11 larger than the reference case based on isolated-ion channel lists. This difference reflects the combined effects of plasma evolution and the reconstruction of NEEC channels. In many interactions of strong laser pulses with solid targets, different initial ionization states may have only a limited influence after rapid ionization. In the present case, however, the moderate laser intensity makes the ionization process and the initial population of free electrons relevant to the subsequent plasma evolution. The different initial conditions and ionization energies corrected for IPD therefore affect the charge state evolution, while the electronic structure at finite density modifies the available NEEC resonances and their overlap with the transient electron spectra. Therefore, treating a dense thorium target as an ensemble of isolated ions may lead to sizable deviations in the estimated NEEC yield and in the charge states that dominate the contribution. This comparison demonstrates that these electronic structure effects should be included when determining NEEC contributions and estimating yields in thorium at solid density.

\section{Conclusion}
\label{sec:conclusion}

We have demonstrated that nuclear excitation by electron capture in dense $^{229}\mathrm{Th}$ is governed by the electronic structure of the surrounding medium and cannot be described by a direct transfer of capture orbitals of isolated ions. Using finite-temperature average-atom calculations, we identify physically accessible NEEC channels through three coupled requirements: electronic localization, resonance energy matching, and vacancy availability. Near metallic density, the shallow $6p$ states remain localized and can support resonant electron capture that drives the 8.356-eV isomeric transition, whereas higher valence-like states become continuum-like and no longer constitute localized capture channels. Calculations at different temperatures and densities further show that the availability of the $6p$ channel is restricted to finite windows in which localization, energetic accessibility, and vacancy formation are simultaneously satisfied. Finite density therefore controls NEEC through both channel filtering and resonance shifts, rather than through a simple shift of binding energies of isolated ions. When these reconstructed channels are used in NEEC postprocessing of 2D3V PIC simulations, the predicted cumulative excitation yield differs by about one order of magnitude from the reference case based on an initially neutral target and isolated-ion channels. This difference reflects the combined effects of electronic structure at finite density on charge state evolution, channel availability, resonance placement, and spectral overlap. Our results establish electronic structure as a key factor controlling pathways for resonant electron capture, underscoring its essential role in nuclear excitation driven by electrons in dense matter. 

\begin{acknowledgments}
	This work was supported by the National Key R\&D Program of China 
	(Grant Nos. 2022YFA1603300 and 2024YFA1613400), the National Natural 
	Science Foundation of China (Grant Nos. 12375244, 12405288, and 12595364), 
	the Hunan Provincial Natural Science Foundation (Grant No. 2026JJ60111), 
	the Hunan Provincial Innovation Foundation for Postgraduate 
	(Grant No. CX20230008), the Innovation Program of Southwestern Institute 
	of Physics (Grant No. 202501WDZCQN011), the Innovation Research Foundation 
	of National University of Defense Technology 
	(Grant Nos. XJQY2024046 and XJJC2024063), the Natural Science Foundation 
	of Top Talent of SZTU (Grant No. GDRC202526), and the Shenzhen Key Laboratory 
	of Ultraintense Laser and Advanced Material Technology 
	(Grant No. ZDSYS20200811143600001).
\end{acknowledgments} 

\section*{AUTHOR DECLARATIONS}

\subsection*{Conflict of Interest}
The authors have no conflicts to disclose.

\subsection*{Author Contributions}

\noindent \textbf{Yang-Yang Xu}: Conceptualization; Methodology; Software; 
Formal analysis; Investigation; Data curation; Visualization; Writing -- original draft; 
Writing -- review \& editing. 
\textbf{Jin-Tao Qi}: Conceptualization; Methodology; Supervision; 
Writing -- review \& editing. 
\textbf{Qiong Xiao}: Methodology; Formal analysis; Investigation; Validation; 
Writing -- review \& editing. 
\textbf{Jun-Hao Cheng}: Methodology; Formal analysis; Investigation; Validation; 
Writing -- review \& editing. 
\textbf{Xin-Yan Li}: Methodology; Supervision; Project administration; 
Funding acquisition; Writing -- review \& editing. 
\textbf{Tai-Wu Huang}: Conceptualization; Methodology; Supervision; 
Project administration; Funding acquisition; Writing -- review \& editing. 
\textbf{Tong-Pu Yu}: Conceptualization; Methodology; Supervision; 
Project administration; Funding acquisition; Writing -- review \& editing.

\section*{Data Availability Statement}
The data that support the findings of this study are available from the corresponding author upon reasonable request.

\section*{REFERENCES}
\bibliography{mybib}

\end{document}